\input{aipcheck}
\newcommand{\fprime}{$f_2^{\prime}(1525)$}

\documentclass[
    ,final            
  ]
  {aipproc}

\layoutstyle{6x9}

\begin{document}

\title{Observation of $K_s^0K_s^0$ resonances \\
in deep inelastic scattering at HERA}

\author{M. Barbi}{
  address={McGill University, 
        Physics Department\\
        Montreal, Quebec\\ 
	Canada}
	\\ (on behalf of the ZEUS Collaboration)
}

\begin{abstract}
 Inclusive $K_s^0K_s^0$ production
in deep inelastic $ep$ scattering at HERA has been studied
with the ZEUS detector 
using an integrated luminosity of $120$~pb$^{-1}$. 
Two states are observed at masses of 
$1537^{+9}_{-8}$~MeV and $1726\pm 7$~MeV, 
as well as an enhancement around $1300$~MeV.
The state at $1537$~MeV is consistent with the well established
\fprime. The state at $1726$~MeV may be the glueball candidate
$f_0(1710)$. However, it's width of $38^{+20}_{-14}$~MeV is narrower 
than $125\pm10$~MeV observed by previous experiments for the $f_0(1710)$.
\end{abstract}

\maketitle


\section{Introduction}

The $K_s^0K_s^0$ system is expected to couple to scalar
and tensor glueballs. This has motivated intense 
experimental and theoretical study during the past few years
\cite{rmp:v71:1411,hep-ex/0101031}.
Lattice QCD calculations \cite{glueball1,glueball2}
predict the existence of a scalar 
glueball with a mass of $1730\pm 100$ MeV and 
a tensor glueball at $2400\pm 120$ MeV.
The scalar glueball can mix with $q\overline{q}$ states with $I=0$ from 
the scalar meson nonet, leading to three $J^{PC}=0^{++}$ states whereas
only two can fit into the nonet.
Experimentally, four states with $J^{PC}=0^{++}$ and $I=0$ have been 
established \cite{pdg}: $f_0(980)$, $f_0(1370)$, $f_0(1500)$ and $f_0(1710)$.

The state most frequently considered to be a glueball candidate 
is $f_0(1710)$ \cite{pdg} , but its gluon content has not 
yet been established. 
This state was first observed in radiative $J/\psi$ 
decays \cite{BES96} and
its angular momentum $J=0$ was established by the WA102 experiment 
using a partial-wave analysis in the $K^+K^-$ and $K_s^0K_s^0$ 
final states\cite{WAf1710} .
Observation of $f_0(1710)$ in $\gamma\gamma$ 
collisions may indicate a large quark content.
A recent publication from L3 \cite{L3} 
reports the observation of two states in $\gamma\gamma$ collisions
above $1500$ MeV, the well-established $f_2^\prime(1525)$ \cite{pdg}
and a broad resonance at 1760 MeV. 
It is not clear if the latter state is the $f_0(1710)$.

The $ep$ collisions at HERA provide an opportunity to study resonance
production in a new environment. Production of $K_s^0$ particles 
has been studied previously at HERA \cite{epj:c2:77,zfp:c68:29,np:b480:3}.
In this contribution, the first observation of resonances in the 
$K_s^0K_s^0$ final state in inclusive deep inelastic $ep$ scattering 
(DIS) is reported \cite{hep-0308006}.

\section{Event selection and $K_s^0$-pair candidates}

A detailed description of the ZEUS detector can be found 
elsewhere~\cite{zeus:1993:bluebook}.

The data used for this study correspond to a total integrated luminosity of 
120 pb$^{-1}$ collected in ZEUS during the 1996-2000 running period. 

The inclusive neutral current DIS process 
$e(k)+p(P) \rightarrow e(k^\prime)+X$
can be 
described in terms of the following variables:
the negative of the invariant-mass 
squared of the exchanged virtual photon,
$Q^2~=~-q^2~=~-$ $(k-k^\prime)^2$; the fraction of the lepton 
energy transferred 
to the proton in the proton rest frame,
$y=(q\cdot P)/(k\cdot P)$;
and the Bjorken scaling variable, 
$x=Q^2/(2P\cdot q)$.

A three-level trigger system was used to select events online 
\cite{zeus:1993:bluebook}.
The inclusive DIS selection was defined by requiring an
electron found in the Uranium Calorimeter, and further requirements were applied
to ensure a well defined data sample~\cite{hep-0308006}.

Oppositely charged track 
pairs reconstructed by the ZEUS central tracking detector (CTD) and 
assigned to a secondary vertex were selected and combined 
to form $K_s^0$ candidates. 
Both tracks were assigned the mass of a charged pion and the 
invariant-mass $M(\pi^+\pi^-)$ was calculated. 
Only events with at least one pair of $K_s^0$ candidates were selected.
The invariant mass of the $K_s^0$ pair candidate $M(K_s^0,K_s^0)$
was reconstructed in the range \mbox{$0.995~<~M(K_s^0K_s^0)~<2.795$ GeV}.
A detailed description of the $K_s^0$ pair candidate selection can be found
in~\cite{hep-0308006}.

\begin{figure}
\includegraphics[height=.4\textheight]{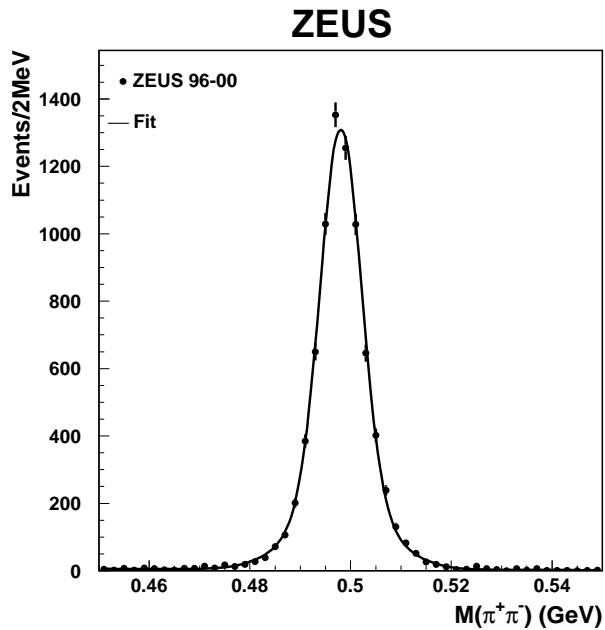}
 \caption{The distribution of $\pi^+\pi^-$ invariant mass 
for events with at least two $K_s^0$ candidates 
passing all selection cuts. The solid line shows the result of a
fit using one linear and two Gaussian functions.}
\label{fig:Ksmass} 
\end{figure}

Figure \ref{fig:Ksmass} shows the distribution in $x$ and $Q^2$ 
of selected events containing at least one 
pair of $K_s^0$ candidates. 

\begin{figure}
\includegraphics[height=.4\textheight]{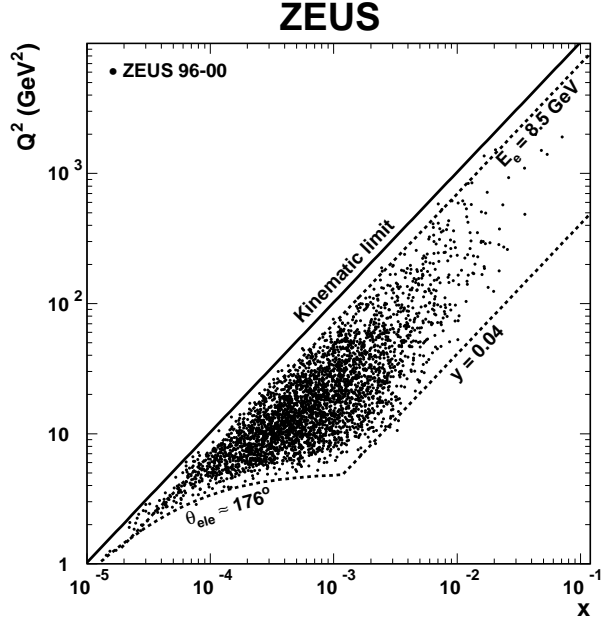}
 \caption{The distribution in $x$ and $Q^2$ of events passing all selection cuts.
The dashed lines delineate approximately the 
kinematic region selected. The solid
line indicates the kinematic limit for HERA running with 920 GeV protons.}
\label{fig:q2vsxbj}
\end{figure}

Figure~\ref{fig:q2vsxbj} shows the $M(\pi^+\pi^-)$ distribution in the range
\mbox{$0.45<M(\pi^+\pi^-)<0.55$ GeV} after the $K_s^0$ pair candidate selection.

\section{Results}

The $K_s^0K_s^0$ spectrum may have a strong enhancement near the
$K_s^0K_s^0$ threshold due to the $f_0(980)$/$a_0(980)$
state \cite{prd:v32:189,pl:b489:24,hep-ph/0202157}.
Since the high $K_s^0K_s^0$ mass 
is the region of interest for this analysis, the complication 
due to the threshold region is avoided by imposing the cut 
\mbox{$cos\theta_{K_s^0K_s^0}<0.92$}, where $\theta_{K_s^0K_s^0}$ 
is the opening angle between the two $K_s^0$ candidates in the 
laboratory frame.

\begin{figure}
  \includegraphics[height=.4\textheight]{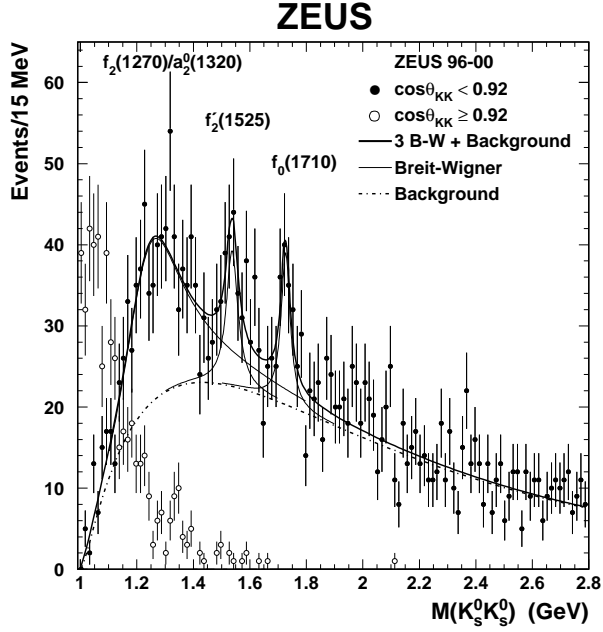}
  \caption{The $K_s^0K_s^0$ invariant-mass spectrum for $K_s^0$ pair candidates with
$cos\theta_{K_s^0K_s^0}<0.92$ (filled circles). 
The thick solid line is the result 
of a fit using three Breit-Wigners (thin solid lines) and a background
function (dotted-dashed line). 
The $K_s^0$ pair candidates that fail the $cos\theta_{K_s^0K_s^0}<0.92$
cut are also shown (open circles).}
\label{fig:KKmass}
\end{figure}

After applying all selections, 2553 $K_s^0$-pair candidates were
found in the range
$0.995<M(K_s^0K_s^0)<2.795$ GeV, where
$M(K_s^0K_s^0)$ was calculated using the $K_s^0$ mass of $497.672~$MeV
\cite{pdg}. The momentum resolution of the CTD leads to an average
$M(K_s^0K_s^0)$ resolution which ranges from $7$ MeV in the 
$1300$ MeV mass region to 10 MeV in the $1700$ MeV region.
Figure \ref{fig:KKmass} shows the measured $K_s^0K_s^0$ invariant-mass 
spectrum.
Two clear peaks are 
seen, one around 1500 MeV and the other around 1700 MeV,
along with an enhancement around 1300 MeV.
The data for \mbox{$\cos{\theta_{K_s^0K_s^0}}>0.92$} are also
shown. 

The distribution of Fig.~\ref{fig:KKmass} was fitted 
using three modified relativistic Breit-Wigner (MRBW) 
distributions 
and a background function $U(M)$;

\begin{equation}
F(M) = \sum_{i=1}^{3}
(\frac{m_{*,i}\Gamma_{d,i}}{(m_{*,i}^2-M^2)^2+m_{*,i}^2\Gamma_{d,i}^2})
+ U(M)~,
\end{equation}

where $\Gamma_{d,i}$ is the effective resonance width, 
which takes into account spin and large width effects \cite{bwigner},
$m_{*,i}$ is the resonance mass, and $M$ is $K^0_sK^0_s$ invariant 
mass. The background function is

\begin{equation}
U(M)~=~A\cdot (M-2m_{K_s^0})^{B}\cdot e^{-C \sqrt{M-2m_{K_s^0}}}~,
\label{bgndfunc}
\end{equation}

where $A$, $B$ and $C$ are free parameters and 
$m_{K_s^0}$ is the $K_s^0$ mass defined by Hagiwara et al \cite{pdg}. 
Monte Carlo studies showed that effects of the track-momentum resolution
on the mass reconstruction were small compared to the measured widths 
of the states. Therefore the resolution effects were ignored in the fit.

Below $1500$ MeV, a region strongly affected by the 
$\cos{\theta_{K_s^0K_s^0}}$ cut, a peak is seen around 
$1300$ MeV where a contribution from $f_2(1270)/a_2^0(1320)$ is expected.
This mass region was fitted with a single Breit-Wigner.

Above $1500$ MeV, the lower-mass state has a fitted mass of 
$1537^{+9}_{-8}$ MeV and a width of $50^{+34}_{-22}$ MeV, in good 
agreement with the well established $f_2^\prime(1525)$.
The higher-mass state has a fitted mass of $1726 \pm 7$ MeV and
a width of $38^{+20}_{-14}$ MeV.  
The widths reported here were stable, within statistical errors,
to a wide variation of fitting methods including those using 30 MeV 
bins rather than the default 15 MeV bins.
The width is 
narrower than the PDG value
($125 \pm10$ MeV) \cite{pdg} reported for $f_0(1710)$, but when 
it is fixed to this value, 
the fit is still acceptable. 

It was found that 93\% of the $K_s^0$-pair candidates selected within the 
detector and trigger acceptance are in the target region of the Breit 
frame \cite{bk:1972:1,zerw:zfp:237}, the hemisphere containing the proton remnant. Of 
the $K_s^0$-pair candidates in the target region, 78\% are in the 
region $x_{p}=2p_B/Q > 1$, 
where $p_B$ is the absolute momentum of the $K_s^0K_s^0$ in the Breit frame.
High $x_{p}$ corresponds to production of the $K_s^0$-pair in a region
where sizeable initial state gluon radiation may be expected. This is in
contrast to the situation at $e^+e^-$ colliders where the particles 
entering the hard scattering are colourless.  

\section{Conclusions}
\label{sec-con}

The first observation in $ep$ deep inelastic scattering of a state 
at $1537$ MeV, consistent with 
$f^\prime_2(1525)$, and another at 1726 MeV, close to $f_0(1710)$, 
is reported. 
There is also an enhancement near 1300 MeV which may arise from the 
production of $f_2(1270)$ and/or $a_2^0(1320)$ states.
The width of the state at $1537$ MeV
is consistent with the PDG value for the $f_2^\prime(1525)$.
The state at 1726 MeV has a mass consistent with the glueball candidate
$f_0(1710)$, and is found in a gluon-rich region of phase space.
However, it's width of $38^{+20}_{-14}$ MeV is narrower than the PDG value 
of 125$\pm$10 MeV for the $f_0(1710)$.



\bibliographystyle{aipproc}

\bibliography{sample}

\IfFileExists{\jobname.bbl}{}
 {\typeout{}
  \typeout{******************************************}
  \typeout{** Please run "bibtex \jobname" to optain}
  \typeout{** the bibliography and then re-run LaTeX}
  \typeout{** twice to fix the references!}
  \typeout{******************************************}
  \typeout{}
 }

\end{document}